# Exploring the effects of diameter and volume fraction of quantum dots on photocarrier generation rate in solar cells

F. Hafiz[1,2], M. R. I. Rafi[1,3], M. Tasfia[1], M. M. Rahman[4] and M. M. Chowdhury[5,6]

[1]Department of Electrical and Electronic Engineering, Bangladesh University of Engineering and Technology, Dhaka-1205, Bangladesh

[2]Department of Computer Science and Engineering, United International University, Dhaka-1212, Bangladesh

[3]Department of Electrical and Electronic Engineering, BRAC University, Dhaka-1212, Bangladesh

[4]School of Electrical and Computer Engineering, Purdue University, West Lafayette, Indiana 47907, United States

[5]Department of Electrical and Computer Engineering, University of British Columbia, Vancouver, British Columbia V6T 1Z4, Canada

[6]Quantum Matter Institute, University of British Columbia, Vancouver, British Columbia V6T 1Z4, Canada

**Abstract**

This paper extends a previous model for *p-i-n* GaAs quantum dot solar cells (QDSC) by revising the equation of photocarrier generation rate in quantum dots (QDs) inside the intrinsic region of the device structure. In our model, we address a notable discrepancy that arose from the previous model where they did not consider the volume of QDs within the intrinsic region, leading to an overestimation of the photocarrier generation rate. Our present model rectifies this by incorporating the volume of quantum dots, resulting in adjustments to the photocarrier generation rate. Additionally, we determine the absorption coefficient of the QDs based on Mie theory for different diameter sizes considering the constant volume fraction of the total number of QDs in the intrinsic region. During this analysis, we find that the absorption spectra of the QDs and host material may overlap in certain cases, although the previous model assumed no overlap. This finding suggests the need for caution when evaluating spectral overlap: if the spectra do not overlap, both the previous and current modified models can be reliably applied. However, in cases of overlap, careful consideration is required to ensure accurate predictions of photocarrier generation. Furthermore, investigating the effect of QD diameter size on the photocarrier generation rate in the intrinsic region, we find that smaller QD sizes result in a higher absorption coefficient as well as a higher generation rate for a constant volume of QDs in the region. Moreover, we establish the optimization of the QDs array size by varying the size as well as the total volume of QDs to improve the generation rate in the solar cell. Our analysis reveals that a higher volume of QDs and a smaller size of QDs result in the maximum generation rate. From an experimental perspective, we propose that the optimal arrangement of QDs in such solar cells is a 0.5 volume fraction with a QD diameter of 2 nm.

## 1. Introduction

In recent years, several new design schemes for solar cells have been proposed to maximize solar energy harnessing[1], [2]. Aqoma et al. [1] proposed an ammonium iodide-based ligand exchange strategy for solar devices for enhanced efficiency. Despite the great progress, the major limitations of conventional solar cell devices are twofold: low-energy photons cannot excite charge carriers to the conduction band, thus not contributing to device current; and high-energy photons are inefficiently utilized due to a poor match with the energy gap [3]. To address these limitations, scientists have proposed various new design schemes such as multi-junction solar cells, quantum well solar cells, etc. [4], [5]. In the study of Aroutiounian et al. [6],a new concept called quantum dot solar cells (QDSCs) for GaAs was introduced to enhance the efficiency of solar energy conversion by utilizing additional photocurrent generated in multi quantum-dot layers stacked in the intrinsic *i*-region of the *p-i-n* structure. QDs possess a zero-dimensional nature with discrete energy levels, making them ideal candidates for intermediate band-based solar cells with a theoretical efficiency of 63% [7]. Although QDSCs have often exhibited improved short-circuit currents compared to bulk single-junction solar cells without QDs, the overall contribution to efficiency enhancement from QDs remains marginal. The concept and analytical modelling of multi-stacked QDs for solar cells was first introduced to increase absorption efficiency in the work of [6]. However, fabricating stacked layers poses challenges due to accumulated strain creating defects and reducing photon absorption [8]. Additionally, achieving higher efficiency in QDSCs requires addressing issues such as QD recombination effects, marginal photocurrent collection, and degradation of open-circuit voltage [9]. In Quantum Well (QW) solar cells, absorption was enhanced by adjusting the position of QW layers to match Fabry-Pérot resonant (FP) peaks [10]. Similarly, optimization of QD layer positions in solar cells resulted in higher absorption and photogeneration rates [11]. While QDSCs face significant challenges, they hold immense potential for becoming the next generation of highly efficient solar cells. Despite theoretical studies on how variations in QD size affect the band gap width of the solar cell and radiative lifetime of the lowest energy state [12], there are few studies on the effect of QD size on absorption and photogeneration rate in the solar cell [14]. While an experimental study [13] has shown the impact of QD height on solar cell performance, further investigation is required to understand how different types and sizes of QDs affect solar cell performance.

In this work, we divide our study into two main parts: (i) proposing an extended model for a *p-i-n* quantum dot (QDs) solar cell, (ii) investigating the effect of spherical QDs size and volume fraction on the absorption coefficient and photogeneration rate of the solar cell as well as proposing optimal QD size and volume fraction. We have identified an overestimation of photocurrent in the study of Aroutiounian et al. [6] and introduced a revised model. Additionally, we have explored the impact of QDs size and volume fraction on absorption coefficient, generation rate, etc., employing Mie theory [14]. Through our investigation, we have determined the optimal size and volume fraction of QDs that yield the maximum generation rate in the solar cell, thereby enhancing the efficiency of a simple *p-i-n* solar cell.

## 2. Methodology

### 2.1 Device Structure and Analytical Modeling

In our study, we have developed a revised ideal model of photocurrent in a *p-i-n* QD solar cell, as depicted in Figure 1. The absorbing material is GaAs, and the QDs consist of InAs material. The device incorporates multi-quantum-dot layers in the intrinsic region to enhance the photocurrent [6]. The intrinsic region is composed of two constituents: the bulk GaAs region within which spherical InAs QDs are stacked, and the bulk GaAs region in the intrinsic region is regarded as the 'barrier region'. The solar cell dimensions were kept the same as in the previous study [6], with the *p*-region width, $x_p = 0.8$ μm; *i*-region width, $x_i = 3$ μm, and the *n*-region width, $x_n = 2$ μm.

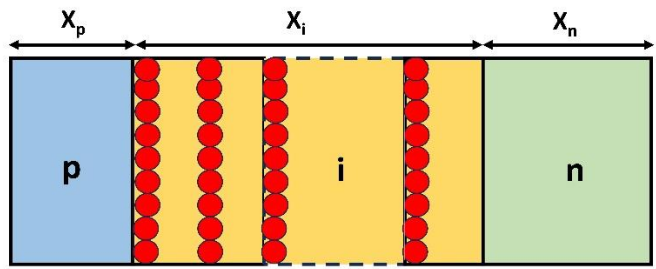

**Figure 1.** Schematic of the QD *p-i-n* solar cell under consideration: The *p*-region, *i*-region, and *n*-region are made of GaAs, while the spherical QDs are made of InAs. We consider volume density, $n_D$ for spherical QDs with a volume of $V_D$ per QD.

Our assumptions include uniform doping in the *p*-type and *n*-type regions, with transport parameters, generation rates, current density being a function of depth ($x$) and incident wavelength ($\lambda$) only, allowing for a one-dimensional transport assumption. We have also assumed quasi-neutrality and low-injection conditions throughout our analytical study.

The transport characteristics are computed with the help of the following equations [15], [16]-

$$\frac{dj_p(x)}{dx} = qG_p(\lambda, x) - q\frac{p(x)-p_0}{\tau_p} \quad \text{and} \quad \frac{dj_n(x)}{dx} = -qG_n(\lambda, x) + q\frac{n(x)-n_0}{\tau_n} \quad (1)$$

Where $j_n$, , $\mu_n$ are the electron current density, mobility respectively. $G_p(\lambda, x)$ and $G_n(\lambda, x)$ are the electron and hole generation rate respectively at a depth, x . for the incident light of wavelength, $\lambda$. Moreover, $q$ is the magnitude of electron's charge and $n(x), p(x)$ represents the total electron and hole concentration respectively at non-equilibrium whereas $n_0, p_0$ are the thermal equilibrium electron and hole concentration respectively. $\tau_n$ and $\tau_p$ are the excess minority carrier electron and hole lifetime respectively. For the incident light of wavelength, $\lambda$ and flux, $F(\lambda)$ the electron-hole generation rate in the *p*-region at a depth *x*, is equal to [6]:

$$G_p(\lambda, x) = F(\lambda)[1 - R(\lambda)]\alpha(\lambda)exp\,[-\alpha(\lambda)x] \quad (2)$$

Now, if (2) is integrated for the wavelength region of GaAs for generation rate in the different positions of the *p*-region, we get:

$$G_P(x) = \int_0^{\lambda_{GaAs}} G_p(\lambda, x)dx = \int_0^{\lambda_{GaAs}} F(\lambda)[1 - R(\lambda)][1 - exp\,(-\alpha(\lambda)x)\,]d\lambda \quad (3)$$

where $\lambda_{GaAs} = 0.9$ µm is the cut-off wavelength of GaAs for absorption [4]. The value of $x$ is varied from 0 to $x_p$ in a linear step manner and at each value of $x$, the integration was performed to calculate the generation rate at different positions in the *p*-region.

Similarly, if (2) is integrated from 0 to $x_p$, then the photon flux absorbed by the QDs per unit area inside the *p*-region can be written according to [6]:

$$\phi_P(\lambda) = \int_0^{x_p} G_P(\lambda, x)dx = F(\lambda)[1 - R(\lambda)][1 - exp(-\alpha(\lambda)x_p)] \quad (4)$$

Furthermore, the photocurrent collected from the p-region can be written as:

$$j_P(\lambda) = q \int_0^{x_p} G_P(\lambda, x)dx = q\phi_P(\lambda) \quad (5)$$

[6], [15], [17][6]

According to Aroutiounian *et al.* [6], the photocarrier generation rate in QDs inside the *i*-region is represented as,

$$G_{QD}(\lambda, x) = F(\lambda)[1 - R(\lambda)]\alpha_{QD}(\lambda) \times exp[-\alpha_{QD}(\lambda)(x - x_p)] \quad (6)$$

where $\alpha_{QD}(\lambda)$ is the QDs' ensemble absorption coefficient. Then the photocurrent collected from the QDs is calculated from,

$$j_{QD}(\lambda) = q \int_{x_p}^{x_p + x_i} G_{QD}(\lambda, x)dx = q\phi_{QD}(\lambda) \quad (7)$$

Now, if (6) is integrated from $x_p$ to $x_p + x_i$, then the photon flux absorbed by QDs per unit area inside the intrinsic region of the device is obtained as follows according to [6],

$$\phi_{QD}(\lambda) = \int_{x_p}^{x_p + x_i} G_{QD}(\lambda, x)dx = F(\lambda)[1 - R(\lambda)][1 - exp(-\alpha_{QD}(\lambda)x_i)] \quad (8)$$

The above equation implicitly refers to the fact that the intrinsic region is filled up only by QDs, there is no empty barrier region available around the QDs, but this cannot be the real picture of a device. Hence the model does not account for the occupied region of barrier region material, GaAs in the intrinsic region, which ultimately leads the model to an overestimation of photocurrent in the QDs.

Therefore, a more appropriate expression to calculate the photocarrier generation rate in QDs inside the *i*-region is proposed in this work,

$$G_{QD}(\lambda, x) = F(\lambda)[1 - R(\lambda)]n_D V_D \alpha_{QD}(\lambda) \times exp\left[-n_D V_D \alpha_{QD}(\lambda)(x - x_p)\right] \quad (9)$$

where $V_D$ is the single QD volume and $n_D$ is the QD volume density. Similarly, QD's generation rate in the different positions of the $i$-region can be written as:

$$G_{QD}(x) = \int_0^{\lambda_{QD}} G_{QD}(\lambda, x)dx = \int_0^{\lambda_{QD}} F(\lambda)[1 - R(\lambda)]\alpha_{QD}(\lambda) \times exp\left[-\alpha_{QD}(\lambda)(x - x_p)\right]d\lambda \quad (10)$$

where $\lambda_{QD} = 3.44$ µm is the cut-off wavelength of InAs QD for absorption [18].

By integrating (9) within the limit of $x_p$ to $x_p + x_i$, the amount of absorbed photon flux within the QDs is obtained as follows:

$$\phi_{QD}(\lambda) = F(\lambda)[1 - R(\lambda)] \times \left[1 - exp\left[-(n_D V_D)\alpha_{QD}(\lambda)x_i\right]\right] \quad (11)$$

On the other hand, to write the generation rate in the barrier region, the presence of QDs in the intrinsic region is taken into account properly in [6]; it is considered that only the fraction $(1 - n_D V_D)$ of the $i$-region is occupied by the GaAs barrier region and the carrier generation rate is expressed as,

$$G_B(\lambda, x) = F(\lambda)[1 - R(\lambda)] \times exp[-\alpha(\lambda)x_p](1 - n_D V_D)\alpha(\lambda) \times$$

$$exp\left[-(1 - n_D V_D)\alpha(\lambda)(x - x_p)\right] \quad (12)$$

The barrier generation rate for different positions in the i-region can be determined by integrating with respect to wavelength till the cut-off wavelength of GaAs, $\lambda_{GaAs}$. Similarly, by integrating (12) within the limit of $x_p$ to $x_p + x_i$, the amount of absorbed photon flux inside the barrier regions is obtained,

$$\phi_B(\lambda) = F(\lambda)[1 - R(\lambda)]exp\left[-\alpha(\lambda)x_p\right] \times \left[1 - exp\left[-(1 - n_D V_D)\alpha(\lambda)x_i\right]\right] \quad (13)$$

Aroutiounian *et al*. [6] showed that the photocurrent collected from the $i$-region, $j_i$ is equal to:

$$J_i = \int_0^{\lambda_1} j_B(\lambda)d\lambda + \int_0^{\lambda_2} j_{QD}(\lambda)d\lambda \quad (14)$$

where $\phi_{QD}(\lambda)$ and $\phi_B(\lambda)$ can be expressed as (11) and (13) directly by inspecting the device structure in Fig. 1.

To obtain the generation rate in the *n*-type region, it should be considered that the photon flux passing through the *n*-type region has already been partially absorbed in the *p*-region and *i*-region of the device structure. Therefore, by accounting for the photon absorption in the *p* and *i*-regions, the expression for generation rate and photon flux in the *n*-type region of the device is as follows:

$$G_N(\lambda, x) = F(\lambda)[1 - R(\lambda)] \times exp\left[-\alpha(\lambda)x_p - (1 - n_D V_D)\alpha(\lambda)x_i\right] \times \alpha(\lambda)exp\left[-\alpha(\lambda)\left(x - (x_p + x_i)\right)\right] \quad (15)$$

$$\phi_N(\lambda) = F(\lambda)[1 - R(\lambda)] \times \exp\left[-\alpha(\lambda)x_p - (1 - n_D V_D)\alpha(\lambda)x_i\right] \times [1 - exp(-\alpha(\lambda)x_n)] \quad (16)$$

Using (15), the photocurrent collected by the *n*-type region is:

$$j_N(\lambda) = q \int_{x_p+x_i}^{x_p+x_i+x_n} G_N(\lambda, x)dx = q\phi_N(\lambda) \quad (17)$$

So, the total photocurrent in each region can be calculated by integrating over the wavelength of the respective material:

$$J_P = \int_0^{\lambda_{GaAs}} j_P(\lambda)d\lambda$$
$$J_i = \int_0^{\lambda_{GaAs}} j_B(\lambda)d\lambda + \int_0^{\lambda_{QD}} j_D(\lambda)d\lambda \quad (18)$$
$$J_N = \int_0^{\lambda_{GaAs}} j_N(\lambda)d\lambda$$

Finally, the overall photocurrent in the device is, $J_{Total} = J_P + J_i + J_N$

In the following section, we discuss how the ensemble QDs absorption coefficient, $\alpha_{QD}(\lambda)$ is calculated with the help of the Mie theory [14].

## 2.2 Mie Theory for QDs Absorption Coefficient Calculation

Although we have information on the wavelength-dependent absorption coefficient of bulk GaAs, determining the absorption coefficient of InAs QDs for different diameters is challenging due to their small size. In order to compare the findings of Aroutiounian et al. [6], we used experimental data for 15 nm InAs QDs. However, we do not have experimental data for when the QD size is varied. To address this, we used Mie theory [14] to calculate the absorption coefficient of the InAs QDs. [14] We have considered the QDs as spherical particles and the absorption while varying the diameter of the QDs can be calculated using the Mie theory. Mie theory is used for determining the absorptions and scattering by particles in various mediums [19], [20], [21]. It is widely used in near-field optics and plasmonics, where light scattering across various particle shapes is studied. Mie theory is very well established for calculating the extinction, scattering, and absorption cross-sections of a spherical particle [22], [23], [24].

If a sphere has a radius $r$ with a refractive index of $n_{sphere}$ and the surrounding medium has a refractive index of $n_{medium}$, the size parameter is defined as $x = kr$, where $k = \frac{2\pi n_{medium}}{\lambda}$ and the relative refractive index is defined as $m = \frac{n_{sphere}}{n_{medium}}$.

According to the Mie theory, to calculate the cross sections, Mie coefficients outside the sphere have to be determined which are:

$$a_n = \frac{\mu m^2 j_n(mx)[x \times j_n(x)]' - \mu_1 j_n(x)[mx \times j_n(mx)]'}{\mu m^2 j_n(mx)\left[x \times h_n^{(1)}(x)\right]' - \mu_1 h_n^{(1)}(x)[mx \times j_n(mx)]'} \quad (19)$$

$$b_n = \frac{\mu_1 j_n(mx)[x \times j_n(x)]' - \mu j_n(x)[mx \times j_n(mx)]'}{\mu_1 j_n(mx)\left[x \times h_n^{(1)}(x)\right]' - \mu h_n^{(1)}(x)[mx \times j_n(mx)]'} \quad (20)$$

Where, $\mu_1$ and $\mu$ are the permeability of the sphere and the medium, respectively; $x$ is the size parameter and integer, $n$ $(1 \leq n \leq x + 4x^{1/3} + 2)$ indicates the order of Bessel functions of the first and second kind.

After calculating these coefficients, the scattering cross-sections can be determined using the following equations:

Scattering cross-sections, $C_{sca} = \frac{2\pi}{k^2} \sum_{n=1}^{n_{max}} (2n+1)(|a_n|^2 + |b_n|^2)$

Extinction cross-sections, $C_{ext} = \frac{2\pi}{k^2} \sum_{n=1}^{n_{max}} (2n+1) Re(|a_n| + |b_n|)$

And absorption cross-sections, $C_{abs} = C_{ext} - C_{sca}$

We have calculated the cross-sections of the QDs using these above equations while considering the GaAs barrier region outside the QDs in the *i*-region. The absorption cross-section represents the effective area (cm$^{-2}$) of the photon, whereas the absorption co-efficient indicates the absorption per unit distance (cm$^{-1}$) of the material. Since the absorption cross-section is calculated for each QD, we can determine the ensemble absorption co-efficient of QDs by simply multiplying the cross-section value with the QDs per volume density, $n_D$ (cm$^{-3}$). This gives the ensemble absorption co-efficient of all the QD in the *i*-region.

While calculating the generation rates in different regions of the solar cell we use analytical equations such as (3), (10), (12), and (15). Material-specific parameters in these equations are either taken from experimental data or calculated using Mie theory (for the absorption coefficient of the QDs).

In previous work, Aroutiounian et al. [6] assumed no overlap in the absorption coefficient spectra between GaAs and InAs QDs within the i-region, and we have also assumed the same in our model. Consequently, both the previous model and our modified model can be reliably applied when there is minimal or no spectral overlap. However, for cases with considerable spectral overlap, careful adjustments in the modeling approach will be necessary to ensure accurate predictions and optimal design of the solar cell.

## 3. Results and discussion

Armed with the equations for carrier generation rates and the expressions for absorption coefficients using Mie theory, we are now ready to calculate the performance of a QD solar cell. In this section, we discuss our overall results and analysis in three segments: i) comparison of QD generation rates with the earlier model [6] in the intrinsic region, which highlights an overestimation in the earlier findings [6]; ii) the impact of varying the diameter and volume fraction of QDs in the solar cell on the generation rate in the *p-i-n* region The parameters used for calculating generation rates in the solar cell are shown in Table 1.

**Table 1.** Parameter Data for Generation Rate Calculation

| Parameters Name | Reference |
|---|---|
| Flux, $F(\lambda)$ | [25] |
| GaAs Reflectance, $R(\lambda)$ | [6] |
| Absorption coefficient of GaAs, $\alpha(\lambda)$ | [26] |
| Absorption coefficient of InAs QD at 15 nm, $\alpha_{QD}(\lambda)$ | [27] |

**3.1 Generation Rates Using Proposed Model and Comparison**

We proposed a revised model for the *p-i-n* solar cell by including an additional parameter volume fraction, $n_d \times V_d$ in the generation rate equations in the *i*-region and *n*-region, as discussed in the previous section[6]. Figure 2a shows the generation rates in the *p*-region, *i*-region, and *n*-region for bulk GaAs and InAs QDs. [6]. It is observable that bulk GaAs contribute to generation in all three regions, while the InAs QDs contribute to generation only in the intrinsic region. The generation rates exhibit an exponentially decreasing trend with position, due to the exponential term in their analytical expression, as discussed in the previous section. In the intrinsic region, we note the overestimation (blue solid line) of the QD generation rate in the model proposed in [6]. According to our proposed model, the QD generation rate has a comparatively lower value (red solid line), as shown in Figure 2a.

In Figure 2b, we present the total contributions by adding the individual contributions of bulk GaAs and InAs QDs in each region. The generation rates in the p-region and n-region are identical to those in Figure 2a, as these regions contain only bulk GaAs. In the intrinsic region, however, we combine the generation rates of bulk GaAs and InAs QDs to obtain the total generation rate. We again observe a discrepancy (black solid line vs. red solid line) between the total generation rate in the intrinsic region for the model in [6] and our proposed model.

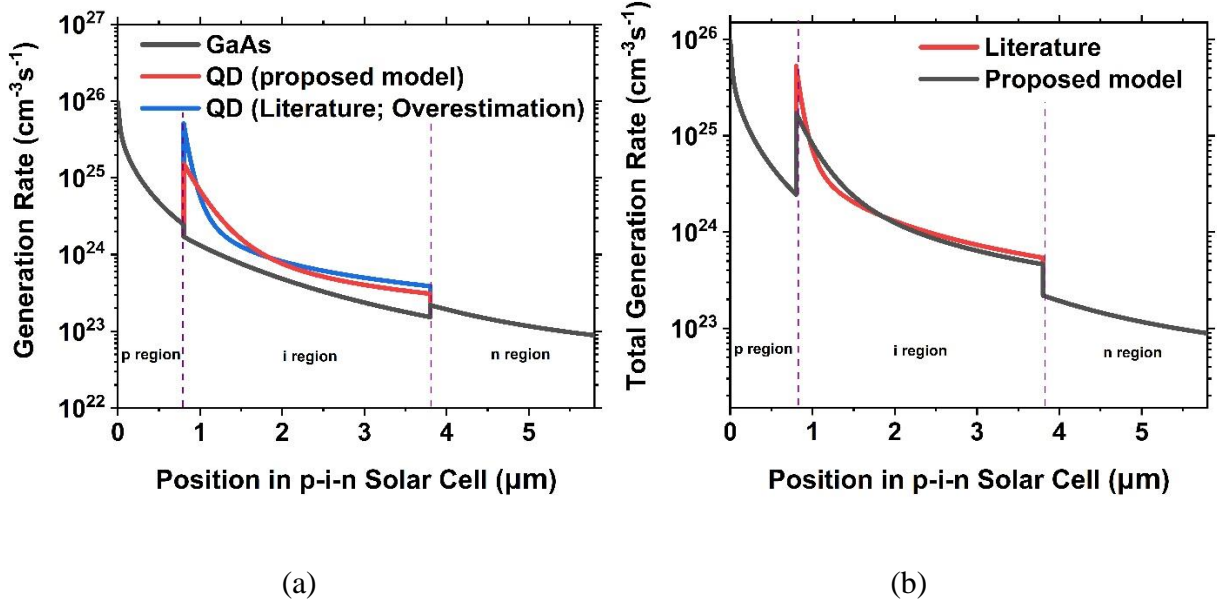

(a)  (b)

**Figure 2. (a)** Generation rates plotted along the QD solar cell (*p*-region, *i*-region, and *n*-region) of the solar cell. The black solid curve represents the generation rates due to bulk GaAs in these regions. The blue and red lines represent the generation rates by the QDs in the intrinsic region for the model proposed in [6] and our proposed model, respectively. **(b)** Total generation rates in different regions for the model in [6] and our proposed model are shown with red and black solid lines, respectively. In Figure 2a, we highlight the contributions of bulk GaAs and InAs QDs to the generation rates in different regions. In Figure 2b, we present the total generation rates across the different regions. The total generation rates in the *p*-region and *n*-region are attributed solely to bulk GaAs, which is why they are identical for both the models in [6] and our proposed model. However, the intrinsic region is the region of interest (ROI), where the overestimation of the generation rates is evident.

### 3.2 Geometrical Effect of QDs on Generation Rate

In this section, we study the geometric effects of ODs on the generation rate of the solar cells. We first explore the effect of changing the QD diameter on the generation rate in all three regions of the *p-i-n* solar cell. Following that, we examine the impact of the QD volume fraction in the *i*-region.

As discussed previously, Mie theory is used to calculate the ensemble absorption coefficient which enabled us to analyze the effect of QD size on various solar cell outputs. Other necessary data used in this numerical calculation are provided in Table 1. The absorption coefficient of the InAs QDs is determined using Mie theory considering the constant volume fraction of the QDs in the barrier region while varying their diameter. Figure 3 shows the absorption coefficient for different sizes of QDs for $n_d \times V_d = 0.1$ and $n_d \times V_d = 0.3$ respectively.

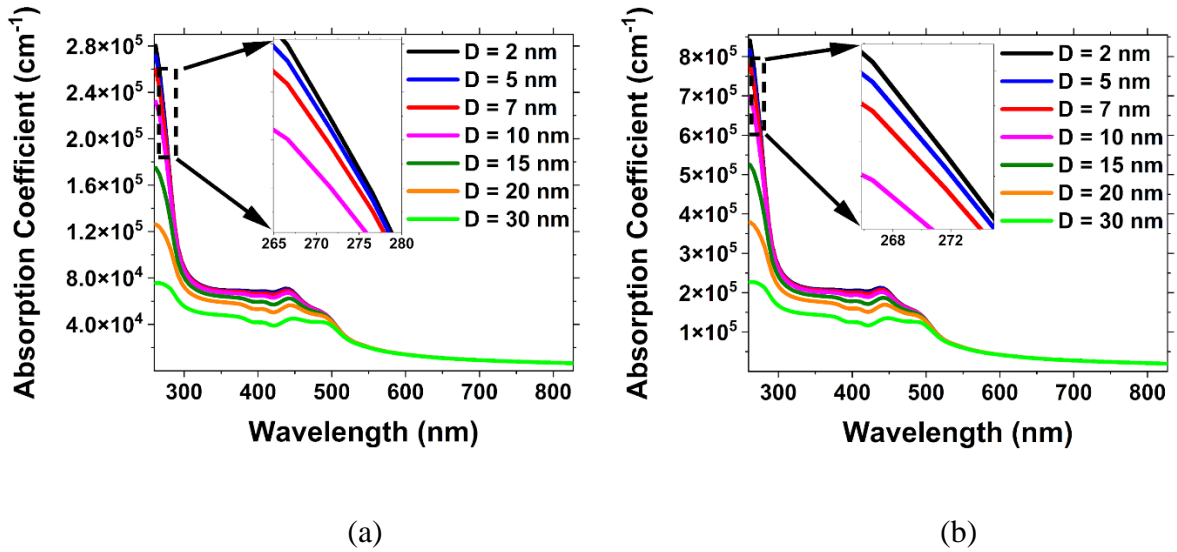

(a)  (b)

**Figure 3.** **(a)** Absorption coefficient, α of InAs QDs for different diameters versus wavelength, λ when $n_d \times V_d = 0.1$. The absorption coefficient decreases by approximately 4 times when the diameter is increased from 2 nm (solid black line) to 30 nm (solid green line). At small diameters, the absorption coefficient values are closely spaced (inset figure) **(b)** Absorption coefficient (cm$^{-1}$) of InAs QDs for different diameters versus wavelength when $n_d \times V_d = 0.3$. We again see that the absorption coefficient decreases approximately 4 times when the diameter is increased from 2 nm (solid black line) to 30 nm (solid green line).

Figure 3 illustrates that the absorption coefficient for 2 nm QDs has a maximum of $2.8 \times 10^5$ cm$^{-1}$ for a constant volume fraction of 0.1 (Figure 3(a)). Conversely, for a constant volume fraction of 0.3, the absorption coefficient for 2nm QDs peaks at $8.5 \times 10^5$ cm$^{-1}$ (Figure 3(b)). This represents an approximately threefold increase in absorption coefficient (solid black line in Figure 3(a) and solid black line in Figure 3(b)) due to an increase in the value of volume fraction, $n_d \times V_d$ from 0.1 to 0.3. Furthermore, as depicted in Figure 3, we observe that the absorption coefficient of InAs QDs increases as the diameter decreases. This phenomenon occurs because the density of QDs increases for smaller QDs when the total volume fraction of QDs is kept constant. Consequently, the overall absorption coefficient is enhanced for smaller QD arrangements in the solar cell. Next, we will use these calculated absorption coefficients to assess the impact on generation rates for different QD sizes and varying volume fractions.

**3.2.1 QD Size Effect on Generation Rate**

Here, the generation rate for the QDs in the intrinsic region is calculated for various diameters of QDs as shown in Figure 4, while keeping the volume fraction of the QDs constant i.e., $n_d \times V_d = 0.1$ and $n_d \times V_d = 0.3$, respectively. The generation rates in the *p*-region and *n*-region are not shown as they remain unchanged when varying the size of the QDs. This is because QD size affects only the absorption coefficient, which in turn influences the generation rate in the intrinsic region alone, as indicated by equations (3) and (18). Figure 4 shows that the generation rate for 2nm QDs peaks at $4.35 \times 10^{24}$ (cm$^{-3}$s$^{-1}$) for a 0.1 volume fraction, while it peaks at $2.1 \times 10^{25}$ (cm$^{-3}$s$^{-1}$) for the 0.3 volume fraction, thus increasing by one order of magnitude (solid black line in Figure 4a and solid black line in Figure 4b) due to an increase in the value of $n_d \times V_d$ from 0.1 to 0.3. Additionally, the inset of Figure 4b illustrates that from the middle to the end of the intrinsic region, the generation rate for larger diameter slightly increases (solid violet line).

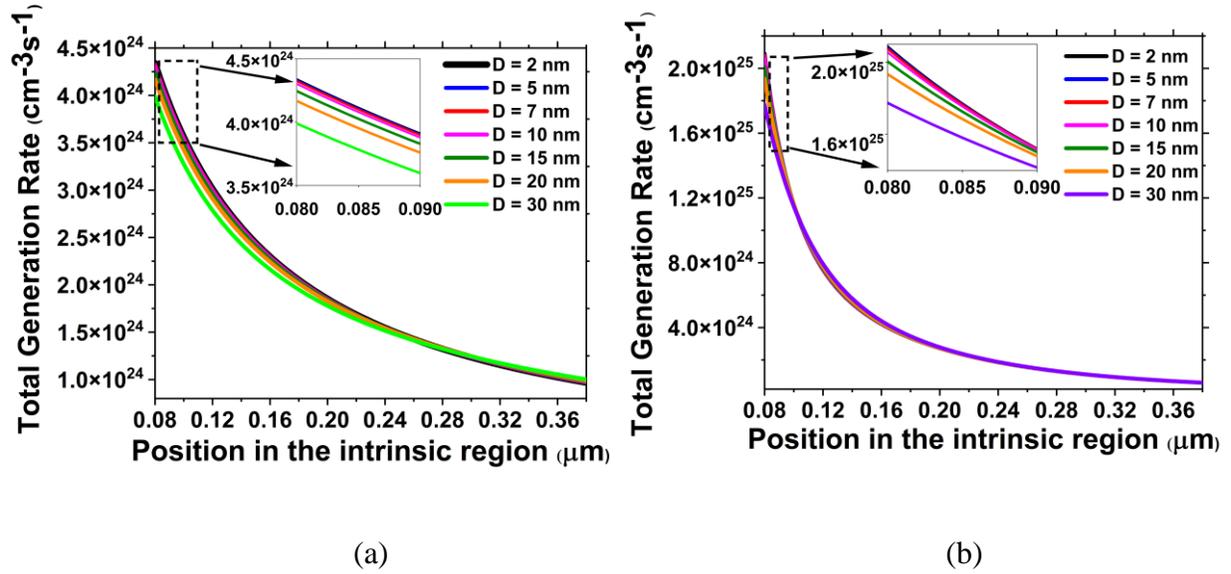

(a)  (b)

**Figure 4.** (**a**) Generation Rate (cm$^{-3}$s$^{-1}$) of InAs QDs for different diameters versus position in the intrinsic region when $n_d \times V_d = 0.1$. The generation rate is higher (solid black line) when 2 nm QDs are incorporated in the intrinsic region. This is because the 2 nm QDs absorb the most incident light as shown in Figure 3 (**b**) Generation Rate (cm$^{-3}$s$^{-1}$) of InAs QDs for different diameters versus position in the intrinsic region when $n_d \times V_d = 0.3$. The generation rate is higher (solid black line) when 2 nm QDs are incorporated in the intrinsic region.

So from the analysis in Figure 4, we find that the generation rate due to the QDs increases as the QDs size decreases

### 3.2.2 QD Volume Fraction Effect on Generation Rate

Next, we turn to studying the effect of QD volume fraction on the generation rate. The generation rate in the *i*-region and *n*-region due to GaAs and QDs also gets affected due to changes in the volume fraction, as we can see from the inclusion of the volume fraction term, $n_d \times V_d$ in equation (9), (11) and (18). However, the generation rate *p*-region is not affected by volume fraction, as indicated by equation (3). This effect of volume fraction on the total generation rate in each region of the solar cell is illustrated in Figure 5.

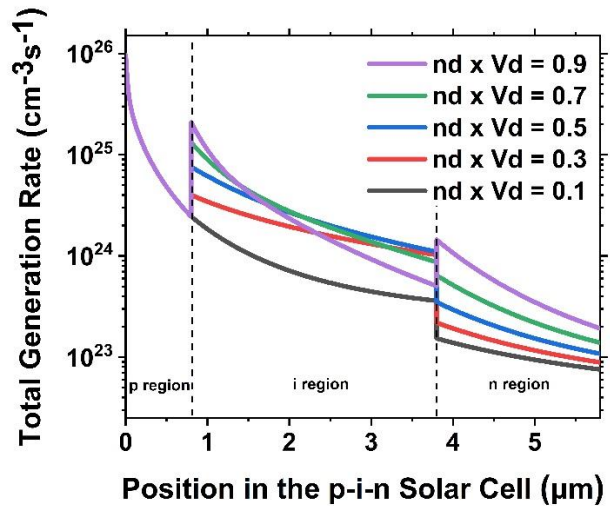

**Figure 5.** Total Generation Rate ($cm^{-3}s^{-1}$) in the *p*-region, *i*-region, and *n*-region for the different volume fractions of the QDs, $n_d \times V_d$. We considered the optimal size of the InAs QD diameter, 2 nm. We observe the generation rate in the *p*-region do not vary with volume fraction.

Figure 5 illustrates an increase in the total generation rate in both the *i*-region and *n*-region as the volume fraction (depicted by the black, red, blue, green, and violet solid lines) of the QDs increases. This increase can be explained by equations (9) and (18). As the volume fraction increases, the QDs absorb more flux from the spectrum, leading to a higher generation rate within the intrinsic region. However, a higher volume fraction of QDs also results in a lower volume fraction of the bulk GaAs, resulting in a lower generation rate by the bulk GaAs in this region. Consequently, the total generation rate in the intrinsic region becomes significantly lower in some cases, such as for a 0.9 volume fraction after the mid-intrinsic region (violet solid line). Similarly, in the *n*-region, a higher volume fraction causes the bulk GaAs in the intrinsic region to absorb less flux, leaving more available flux for the bulk GaAs. This is why higher volume fractions of QDs correspond to increased generation rates by the bulk GaAs in the *n*-region. Although Figure 5 clearly shows that a higher volume fraction of the QDs results in a higher generation rate in the *n*-region, the same cannot be said for the *i*-region. In the *i*-region, the generation rate can be higher for either 0.5, 0.7, or 0.9 volume fractions, indicating that further studies are needed to find the optimal volume fraction. We calculate the total current density by integrating the generation rates from Figure 5 to determine the optimal volume fraction for higher generation rates. Figure 6 shows the bar chart of the current density for different volume fractions.

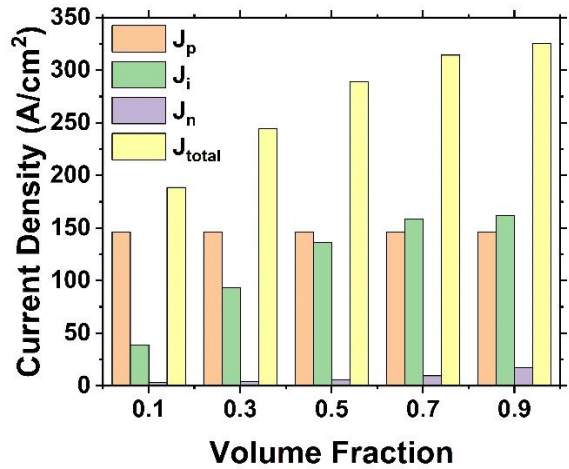

**Figure 6.** Bar chart of Current density (A/cm$^2$) versus volume fraction for the InAs/GaAs solar Cell. We observe that the total current density (yellow-colored bar) increases with volume fraction.

In Figure 6, we observe that the current density for the *i*-region (blue-colored bar) and *n*-region (orange-colored bar) increases as the QD volume fraction increases. As a result, the total current density (yellow-colored bar) also increases. This indicates that, theoretically, higher generation rates and current density can be achieved with a higher volume fraction where the maximum QD volume fraction can be up to 1.

Our geometrical effect analysis concludes that smaller QDs and high-volume fractions can result in an overall higher generation rate in a QD-based solar cell. Experimentally, a QD size of 2 nm and a volume fraction of 0.5 appear suitable for achieving higher generation rates in the InAs/GaAs *p-i-n* solar cell.

## 4. Conclusion

We propose an improved model for photocarrier generation in quantum dot (QD)-based *p-i-n* solar cells by incorporating both the QD and barrier materials, addressing the overestimation of the generation rate in earlier models [6]. The modified model shows that the QD generation rate is significantly lower than previously predicted, which could lead to inaccurate I-V characteristics and efficiency estimates. The absorption coefficients of the QDs were calculated using Mie theory, assuming a spherical QD geometry. We further examine the effects of QD diameter and volume fraction in the intrinsic region on the photogeneration rate. Smaller QDs result in a higher absorption coefficient and photogeneration rate while increasing the QD volume fraction enhances the current density. The study also highlights the importance of considering the spectral overlap between QDs and host materials. Our study provides insights into optimizing QD solar cells by refining the modeling of photogeneration rates and highlighting key geometric factors.

Furthermore, we observed that the absorption spectra of the QDs and host material may overlap in certain cases during this study, even though the previous model assumed no such overlap [6]. This finding suggests a need for caution when evaluating spectral overlap. If no overlap exists, both the prior and current modified models can be used; however, if overlap does occur, a more careful analysis is required to ensure accurate photocarrier generation predictions.

While our detailed optimization approach from a geometric perspective should enhance the efficiency of the solar cell, it is important to note that optimizing the absorption coefficient and generation rates are not the sole determinants of efficiency in building solar cells. While these factors undoubtedly contribute to the likelihood of building efficient solar cells, further theoretical and experimental research is necessary to identify the most suitable geometry for efficiently harnessing solar energy through the development of this novel QD solar cell.